\documentclass[conference]{IEEEtran}

\usepackage{amsmath}
\usepackage{amssymb} 

\usepackage[labelformat=simple]{subcaption}
\DeclareCaptionLabelSeparator{periodspace}{.\quad}
\usepackage{tikz}
\usetikzlibrary{arrows,matrix,positioning,patterns}
\usetikzlibrary{calc}
\usepackage{pgfplots} 
\usetikzlibrary{plotmarks}

\usepackage[utf8]{inputenc}
\usepackage{pgfplots}
\usepgfplotslibrary{groupplots}

\usepackage{url}
\usepackage{multirow}
\usepackage{colortbl}

\usepackage{algorithm} 
\usepackage{algpseudocode}

\DeclareSymbolFont{bbold}{U}{bbold}{m}{n}
\DeclareSymbolFontAlphabet{\mathbbold}{bbold}

\usepackage{listings}
\usepackage{xcolor}
\usepackage[numbers,sort&compress]{natbib}
\usepackage{hyperref}

\IEEEoverridecommandlockouts                              

\title{\LARGE \bf
Simple Trajectory Smoothing for UAV Reference Path Planning \\Based on Decoupling, Spatial Modeling and Linear Programming}
\author{Mogens Plessen*
\thanks{*{\tt\small pmogens@proton.me}, Findklein GmbH,  Switzerland}
}
\renewcommand{\baselinestretch}{0.90}
\begin{document}

\maketitle
\thispagestyle{empty}
\pagestyle{empty}

%
%
%
%


\begin{abstract}
A method for trajectory smoothing for UAV reference path planning is presented. It is derived based on the dynamics of a Dubins airplane model, and involves a decoupling step, spatial modeling and linear programming. The decoupling step enables algebraic control laws for flight-path angle and speed control. Only for roll angle control an optimization step is applied, involving the solution of a small linear program. Two variations are discussed. They differ by reference centerline tracking and the introduction of a path shaping constraint. The benefit of natural dimensionality reduction for spatial modeling is discussed. The simplicity of the overall method is highlighted. An extension to aerobatic flight is outlined, which comes at the cost of a model approximation, however at the gain of maintaining the general model structure. An extension of the method to tractor path planning along 3D terrain is discussed. The method is validated in simulations.
\end{abstract}

\begin{IEEEkeywords}
UAV path planning, trajectory smoothing, Dubins vehicle, spatial modeling, linear programming. 
\end{IEEEkeywords}


\section{Introduction\label{sec_intro}}

In 2D, a Dubins path \cite{dubins1957curves, zhou2026seventy} connects a start and end pose (position and heading) along the shortest possible path. This path is composed of 2 arcs of minimimum turning radius of the vehicle and a straight-line segment. In 3D, for the equivalent problem path planning for Uncrewed Aerial Vehicles (UAVs) is more complex. Depending on the start and end pose there may be an infinite number of paths with minimum pathlength.

Connecting a start and end pose along the shortest possible path is useful in many scenarios. It comes, however, at two costs. First, motion is constrained to either a maximal sharp left turn, a maximal sharp right turn or straight driving. Second, the simplifying assumption is made that paths with instantaneous change from a maximal sharp left turn to a maximal sharp right turn and vice versa are possible.

In many scenarios paths are desired that do not only operate at steering limits and instead may satisfy other constraints. The development of a method for this purpose provided the motivation for this article. Starting from the \emph{Dubins airplane model} proposed in \cite{mclain2014implementing}, an extensive search of all its citing work revealed a research gap for an UAV reference path planning technique that exploits the special structure of the  Dubins airplane model for controls decoupling, spatial modeling and linear programming. This is the contribution of this paper.

A literature review is provided. First, the term 'Dubins airplane model' is not unique. For example, the Dubins airplane models in \cite{chitsaz2007time,lim2026autonomous,mclain2014implementing} are all different. The latter reference relates heading angle to roll angle. Its model includes 3D position coordinates as well as airspeed, flight-path, heading and roll angle. It is therefore our preferred model. 

Starting from a chosen Dubins airplane model a plethora of model-based control algorithms can be applied \cite{kumar2025comprehensive}. This paper paper focuses on an optimization-based approach, which can account for constraints in a structured manner \cite{kamel2017model}. For this purpose, the simplest class of convex optimization problems is linear programming.

\begin{figure}
\centering
\resizebox{\linewidth}{!}{
\begin{tikzpicture}
\draw[black,->, >=latex'] (0,0) -- (4,1);
\node[color=black] (a) at (4.2, 1.1) {$x$};
\draw[black,->, >=latex'] (0,0) -- (-2,1.5);
\node[color=black] (a) at (-1.8, 1.7) {$y$};
\draw[black,->, >=latex'] (0,0) -- (0,2);
\node[color=black] (a) at (-0.05, 2.2) {$z$};
\draw[fill=gray, fill opacity=0.1,blue,thick] (0.7,1.2) -- (1.4,1.15) -- (1.8,2.1) -- (0.7,1.2);
\draw[blue,dashed] (1.4,1.15) -- (1.7,1.125);
\draw[black,dashed] (1.05,1.175) -- (1.5,0.9);
\draw[->,>=latex',red] ($ (1.05,1.175) + 0.5*({cos(-30)},{sin(-30)})$) arc (-30:-3:0.5);
\node[color=red,thick] (a) at (1.7, 0.9) {$\phi$};
\draw[black,dashed] (1.8,2.1) -- (2.5,2.26);
\draw[green,thick,->, >=latex'] (1.8,2.1) -- (2.3,2.7);
\draw[->,>=latex',red] ($ (1.8,2.1) + 0.4*({cos(11)},{sin(11)})$) arc (11:50:0.4);
\node[color=red,thick] (a) at (2.35, 2.4) {$\gamma$};
\node[color=green,thick] (a) at (1.95, 2.7) {$v$};
\draw[black,dashed] (1.05,1.175) -- (3.3,1.7);
\draw[blue,dashed] (1.05,1.175) -- (2.65,1.92);
\draw[blue,dotted] (1.8,2.1) -- (1.8,1.55);
\draw[->,>=latex',red] ($ (2.22,1.43)$) arc (20:90:0.26);
\node[color=red,thick] (a) at (2.4, 1.6) {$\psi$};
\end{tikzpicture}
}
\caption{Illustration of the Dubins airplane model. For simplicity it is depicted as a triangle. States and control variables are annotated. Different colors are used to differentiate positions, angles and velocity. See \eqref{eq_dubins} for interpretation.}
\label{fig_dubinsairpl}
\end{figure}
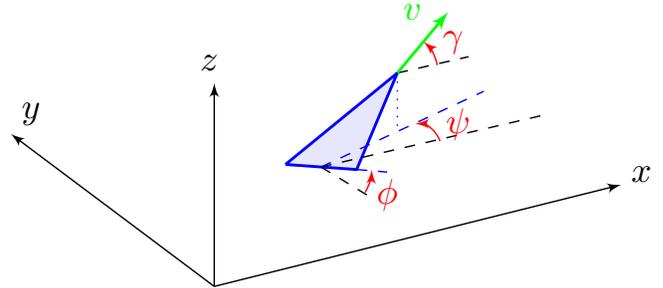

In \cite{vavna2020minimal} a method for 3D Dubins path planning with bounded curvature and pitch angle is presented. The problem is decoupled into horizonal and vertical problems, before both are solved separately using 2D Dubins paths. The focus is exclusively on Dubins paths between start and end pose. The method is not generalizable to trajectory smoothing tasks. Their Dubins airplane model does not include roll angle.

Transforming time-dependent to spatially-dependent dynamics, typically in combination with respect to a reference trajectory, is a technique for problem reformulations \cite{sampei1991path,micaelli1993trajectory}. It was found to often facilitate problem solutions to account for spatial constraints, and is also the approach taken here.

In 2D for planar path planning, the spatial problem reformulation is often referred to as planning in the \emph{Frenet frame} \cite{do2016differential, werling2010optimal}. An extension to 3D is referred to as the \emph{Frenet-Serret} frame, however, there exist more and modeling is much more complex \cite{arrizabalaga2024universal, fork2023euclidean,ramirez2021gravity}.

Within a typical hierarchical framework, the presented method follows after an upstream planning step. For example, given a first high-level routing result \cite{yakunin2025algorithm, plessen2025path}, a second trajectory smoothing step is required to prepare a reference path plan, before a subsequent third low-level path tracking step follows. Thus, the method can be classified as a trajectory smoothing step in-between routing and low-level tracking.

The remaining paper is organized as follows: problem formulation, problem solution, numerical examples, discussion and the conclusion are described in Sect. \ref{sec_intro}-\ref{sec_conclusion}.

\section{Problem Formulation\label{sec_probformul}}

The addressed problem can be formulated as follows. Given an initial (rough) path trajectory in 3D, $\{x_i^\text{ref},y_i^\text{ref},z_i^\text{ref}\}_{i=0}^{N_{\text{ref}}}$, it is sought to \emph{smooth} this trajectory subject to (i) UAV dynamics according to a Dubins airplane model, and (ii) additional trajectory shaping constraints. A desired result is a smoothed reference path plan that improves upon the initial trajectory.

The initial path trajectory may consist of a sequence of waypoints or an interpolated path connecting multiple waypoints. The path may initially not be feasible with respect to UAV dynamics.

The \emph{Dubins airplane model} used for the remainder shall be described by the following set of differential equations \cite{mclain2014implementing},
\begin{align}
\begin{bmatrix} \dot{x}\\ \dot{y} \\ \dot{z} \\ \dot{\psi} \end{bmatrix} &= \begin{bmatrix} v \cos(\gamma) \cos(\psi) \\ v \cos(\gamma) \sin(\psi) \\ v \sin(\gamma) \\  \frac{g}{v}\tan(\phi) \end{bmatrix},\label{eq_dubins}
\end{align}
where $(x,y,z)$ denote the intertial position of the UAV. The remaining variables are airspeed $v$, yaw (heading) angle $\psi$, flight-path angle $\gamma$, roll angle $\phi$, and $g$ the acceleration due to gravity. See Fig. \ref{fig_dubinsairpl} for illustration.

Several comments can be made. First, the Dubins airplane model is a simplified model. It does not account for effects such as aerodynamics, wind, sideslip and transients resulting from low-level control delays. However, for high-level path planning it is well suited \cite{mclain2014implementing}.

Second, it is noted that there exist a variety of alternative models, that in their respective papers are also referred to as \emph{Dubins airplane model} \cite{chitsaz2007time,lim2026autonomous,lin2014path}. Thus, \eqref{eq_dubins} does not represent \emph{the} Dubins airplane model. However, its formulation appears favorable since it involves all 6 variables typical for UAV path planning: 3 inertial position coordinates, yaw, roll and flight-path angle. This serves well for fixed-wing path planning, but in general likewise also for quadrotor path planning \cite{lin2014path}. 
 
Finally, it is explicitly stressed that the topic of this paper is \emph{reference} path planning. Low-level control (such as, e.g., necessary for weather-dependent wind rejection) is not a topic of this paper. Generating smoothed nominal and deterministic reference path plans, that account for nonholonomic robot dynamics in 3D (Dubins airplane model) and spatial precision constraints is important for high-level mission planning. Agricultural area coverage is an example application \cite{plessen2025path}. 

\section{Problem Solution\label{sec_probsoln}}

The proposed solution to the problem of Sect. \ref{sec_probformul} is centered around the Dubins airplane model \eqref{eq_dubins} and exploiting its special structure. There are 4 state variables, $(x,y,z,\psi)$, and 3 control variables, $(\gamma,\phi,v)$. Control methods for each of them are derived in the following.

\subsection{Decoupling of dynamics}

A first observation with respect to \eqref{eq_dubins} is made. While $(x,y)$ are coupled with respect to yaw angle $\psi$, this is not the case for $z$-channel dynamics. This motivates a decoupling. 

Therefore, the following hierarchical control approach is taken: (i) $z$-dynamics are decoupled from the remaining system and treated isolatedly to determine a control law for $\gamma$, before (ii) feeding the resulting $\gamma$-trajectory as a reference to the remaining 3-state dynamics. These result in 
\begin{align}
\begin{bmatrix} \dot{x}\\ \dot{y} \\ \dot{\psi} \end{bmatrix} &= \begin{bmatrix} \bar{v} \cos(\psi) \\ \bar{v} \sin(\psi)  \\  \frac{g}{v}\tan(\phi) \end{bmatrix},\label{eq_3statedynamics}
\end{align}
where $\bar{v}=v\cos(\bar{\gamma})$, and $\bar{\gamma}$ denotes the control trajectory determined from step (i) after decoupling.

\subsection{Step 1: Control of flight-path angle $\gamma$\label{subsec_gamma}}

A control law for flight-path angle $\gamma$ is derived. Integrating the decoupled $z$-dynamics and discretizing under a zero-order hold assumption, $\gamma(t)=\gamma_k,~\forall t\in[kT_S,(k+1)Ts]$, and sampling time $T_s$, one obtains $z_{k+1}=z_k+T_s v_k \sin(\gamma_k)$. Abbreviating sampling space,
\begin{equation}
D_{s,k}=T_sv_k,\label{eq_Dsk}
\end{equation}
accounting for an appropriately interpolated given initial reference trajectory, $\{z_k^\text{ref}\}_{k=0}^N$, along $N+1$ points, and after inversion, the basic control law becomes
\begin{equation}
\gamma_{k}^{\text{ref}}=\text{arcsin}(\frac{z_\text{k+1}^\text{ref} - z_k^\text{ref}}{D_{s,k}^\text{ref}}).
%
\end{equation}
Two refinement steps are additionally applied. First, rate and limit constraints are enforced. Thus, $\tilde{\gamma}_0=\gamma_0^\text{ref}$ and for subsequence indices $k>0$,
\begin{align}
l_{\gamma,k} &= \text{max}(\gamma_\text{min},\tilde{\gamma}_{k-1}-\Delta \gamma_\text{min}),\label{eq_lgamma}\\
u_{\gamma,k} &= \text{min}(\gamma_\text{max},\tilde{\gamma}_{k-1}+\Delta \gamma_\text{max}),\\
\tilde{\gamma}_k &= \text{min}(\text{max}(\gamma_k^\text{ref},l_{\gamma,k}),u_{\gamma,k}),
\end{align}
where $\Delta \gamma_\text{min}=D_{s,k}|\dot{\gamma}_\text{min}|/v_k$, $\Delta \gamma_\text{max}=D_{s,k}\dot{\gamma}_\text{max}/v_k$, and $\dot{\gamma}_\text{min}$, $\dot{\gamma}_\text{max}$, $\gamma_\text{min}$ and $\gamma_\text{max}$ describe rate and limit bounds, respectively. Second, a simple heuristic prediction compensation step is applied, 
\begin{equation}
\bar{\gamma}_k=(\tilde{\gamma}_k+\tilde{\gamma}_{k+1})/2,~\forall k=0,\dots,N-1,\label{eq_bargamma}
\end{equation}
which improved reference tracking. 

Several comments can be made. First, the control law is defined \emph{spatially} along the reference trajectory. This resulted naturally, as a consequence of the specific dynamics, $\dot{z}=v\sin(\gamma)$, and above zero-order hold assumptions used for derivation.

Second, it is noted that rate and limit bounds are in general operating point varying, e.g., as a function of altitude. This can directly be accounted for in above formulation.

Finally, it is noted that the resulting control trajectory $\{\bar{\gamma}_k\}_{k=0}^{N-1}$ according to above formulation is clearly strongly depending on the reference trajectory, $\{z_k^\text{ref}\}_{k=0}^{N}$. However, this is a natural result, compatible with the special dynamics structure of \eqref{eq_dubins}.

\subsection{Step 2: Control of roll angle $\phi$ \label{subsec_phi}}

Control of roll angle is based on the decoupled dynamics of equation \eqref{eq_3statedynamics} with 3 remaining states. Two observations can be made. First, \eqref{eq_3statedynamics} closely resembles the classic kinematic bicycle model for 2D path planning, $[\dot{x}_\text{car},\dot{y}_\text{car},\dot{\psi}_\text{car}]=[v_\text{car} \cos(\psi_\text{car}), v_\text{car}\sin(\psi_\text{car}), v_\text{car} \tan(\delta_\text{car})/l_\text{car}]$, where $\delta_\text{car}$ models steering angle, the center of gravity of the vehicle is assumed to be located at the rear acles, $l_\text{car}$ denotes the vehicle's wheelbase, and subindex $\{.\}_\text{car}$ is used to clearly distinguish the 2 models. This kindematic model served as the basis for a spatial control law for ground-based path smoothing for an agricultural application \cite{plessen2025smoothing}.

Second, it is noted that the control $\{\bar{\gamma}_k\}_{k=0}^{N-1}$ according to derivation in Sect. \ref{subsec_gamma} is naturally spatially dependent.

Both observations motivate to derive a spatially varying control law for $\phi$. For notation, for an arbitrary variable, $p\in\mathbb{R}$, let time and spatial derivatives be denoted by $\dot{p}=\frac{dp}{dt}$ and $p'=\frac{dp}{ds}$, respectively, where $s$ represents the spatial coordinate along a reference trajectory. Coordinates between the time and spatial domain can be transformed by projecting point masses (such as e.g. the vehicle's center of gravity) to the reference trajectory. Let heading deviation be $\dot{e}_\psi=\dot{\psi}-\dot{\psi}_s^\text{ref}$ where $\psi_s^\text{ref}$ is the reference heading, $\dot{e}_y=\bar{v}\sin(e_\psi)$ denoting lateral deviation, and $\dot{s}=\frac{\rho_s\bar{v}\cos(e_\psi)}{\rho_s-e_y}$ where $\rho_s$ denotes the radius of curvature. With $e_\psi' = \frac{\dot{e}_\psi}{\dot{s}}$ and $e_y'=\frac{\dot{e}_y}{\dot{s}}$, the spatial equivalent to \eqref{eq_3statedynamics} is derived as
\begin{align}
\begin{bmatrix} e_\psi' \\ e_y' \end{bmatrix} &= \begin{bmatrix} \frac{(\rho_s-e_y)g\tan(\phi)}{\rho_sv\bar{v}\cos(e_\psi)} - \psi_s'  \\  \frac{(\rho_s-e_y)\tan(e_\psi)}{\rho_s} \end{bmatrix}.\label{eq_spatial}
\end{align}
A key difference with respect to its ground-based equivalent is the velocity-dependence in the denominator of the $e_\psi$-dynamics in \eqref{eq_spatial}. 

Note the dimensionality reduction due to the spatial transformation. In \eqref{eq_spatial} there are only 2 states in comparison to 3 for its time-based equivalent in \eqref{eq_3statedynamics}. Dimensionality reduction is an important characteristic and advantage of spatial modeling.

While the control trajectory $\{\bar{\gamma}_k\}_{k=0}^{N-1}$ from Sect. \ref{subsec_gamma} can be seen as a passive local filtering of $\{z_k^\text{ref}\}_{k=0}^{N}$, for the control of $\phi$ an active optimization-based approach is proposed that determines an optimal solution accounting for the entire horizon of the reference trajectory ahead. 

In line with the problem formulation in Sect. \ref{sec_probformul} and in reference to \cite{plessen2025smoothing}, two linear programs (LP) are proposed: one for reference centerline tracking, and one to illustrate how above spatial modeling permits to naturally formulate spatial constraints.

After linearization and discretization of \eqref{eq_3statedynamics}, the proposed first LP for reference tracking, referred to as \textsf{LP1} in the following, is: 
\begin{subequations}
\label{eq_LP1}
\begin{align}
\min\limits_{ \{ \phi_k \}_{k=0}^{N-1}} &\ \  \sum\nolimits_{j=k}^N |e_{y,k}- e_{y,k}^\text{ref} | \label{eq_LP1_objFcn}\\
\mathrm{s.t.} &\ \ D_{s,k} \frac{\dot{\phi}_\text{min}}{v_k^\text{ref}}  \leq \phi_{k+1} - \phi_k \leq D_{s,k} \frac{\dot{\phi}_\text{max}}{v_k^\text{ref}},\nonumber\\
&   \hspace*{3.5cm} \ k = 0,\dots,N-2,\\ 
&\ \ \phi_\text{min} \leq  \phi_k \leq \phi_\text{max},  \ k = 0,\dots,N-1,
\end{align}
\end{subequations}
with $D_{s,k}=s_{k+1}-s_k,\forall k = 0,\dots,N-2$.

The proposed second LP, referred to as \textsf{LP2} in the following, is: 
\begin{subequations}
\label{eq_LP2}
\begin{align}
\min\limits_{ \{ \phi_k \}_{k=0}^{N-1}} &\ \  \sum\nolimits_{k=1}^N |e_{y,k}- e_{y,k}^\text{ref} | \label{eq_LP2_objFcn}\\
\mathrm{s.t.} &\  \ e_{y,k} - e_{y,k}^\text{ref} \leq 0, \ k = 1,\dots,N, \label{eq_LP2_eymax_cstrts}\\
&\ \ D_{s,k} \frac{\dot{\phi}_\text{min}}{v_k^\text{ref}}  \leq \phi_{k+1} - \phi_k \leq D_{s,j} \frac{\dot{\phi}_\text{max}}{v_k^\text{ref}},\nonumber\\
&  \hspace*{3.5cm} \ k = 0,\dots,N-2,\label{eq_LP2_ddelta_cstrts}\\ 
&\ \ \phi_\text{min} \leq  \phi_k \leq \phi_\text{max},  \ k = 0,\dots,N-1,\label{eq_LP2_delta_cstrts}
\end{align}
\end{subequations}
with $D_{s,k}=s_{j+k}-s_k,\forall k = 0,\dots,N-2$. The key novelty with respect to \eqref{eq_LP1} is the introduction of constraint \eqref{eq_LP2_eymax_cstrts}. This enforces the resulting smoothed trajectory to laterally rest to the right-hand side of the reference path within the projected $(s,e_y)$-plane. Many variations of this constraint are possible, for example, for obstacle avoidance or for other path shaping objectives. 

Note that in both \eqref{eq_LP1} and \eqref{eq_LP2}, speed $v$  is not treated as a control variable. Instead, it is treated as a reference parameter, $\{v_k^\text{ref}\}_{k=0}^N$, that is in general spatially varying along the reference path. This procedure ensures compatibility with $\{\bar{\gamma}_k\}_{k=0}^{N-1}$-control generation according to Sect. \ref{subsec_gamma}. Speed control, hierarchically following both the flight-path angle control and the LP-solution, is discussed below in Sect. \ref{subsec_speedctrl}.

For linearization and in the objective functions, references $e_{\psi,k}^\text{ref}$, $e_{y,k}^\text{ref}$ and $\phi_k^\text{ref}$ are initialized at zero. For the absolute values in \eqref{eq_LP1_objFcn} and \eqref{eq_LP2_objFcn}, $N$ non-negative surrogate variables are introduced and the set of inequality constraints is extended accordingly. For feasibility guarantee, constraints in \eqref{eq_LP2_eymax_cstrts} are softened to, $e_{y,k} - e_{y,k}^\text{ref} \leq \sigma $, by introducing a non-negative slack variable $\sigma\geq 0$ and adding it to the objective function \eqref{eq_LP1_objFcn} with a high weight such that $\sum\nolimits_{k=1}^N |e_{y,k}- e_{y,k}^\text{ref} |$ is replaced with $\sum\nolimits_{k=1}^N |e_{y,k}- e_{y,k}^\text{ref} | + 10^{16}\sigma$. Thus, in total there are $n_u=2N$ and $n_u=2N+1$ scalar real-valued optimization variables for \textsf{LP1} and \textsf{LP2}, respectively. This implies very small LPs, favorable for fast LP-solution times and enabling planning over large horizons. One additional benefit of proposed LP-formulations is that they are hyperparameter-free. There are no weighting parameters in the objective functions.

In general, \emph{LP-iterations} or refinement steps can be conducted. These are useful for heuristic reduction of jaggedness \cite{plessen2025smoothing}. For a first LP the original reference path is used as a reference. After that, the LP-solution itself is used as reference for a second LP-iteration, and so forth. In practice, one refinement step was found to be sufficient.

\subsection{Step 3: Speed control $v$\label{subsec_speedctrl}}

The third and remaining control variable is speed $v$. According to above methodology there are 2 decouplings. First, flight-path angle $\gamma$ is controlled decoupledly from the other 2 control variables according to Sect. \ref{subsec_gamma}. It is determined as a function of $\{z_k^\text{ref}\}_{k=0}^{N}$, but implicitly also as a function of speed via \eqref{eq_Dsk}, which determines the spatial interpolation spacing based on which the spatial control law for $\{\bar{\gamma}_k\}_{k=0}^{N-1}$ is derived. Second, $\{\phi_k\}_{k=0}^{N-1}$ is determined according to the LP-method from Sect. \ref{subsec_phi}, and based on the same spatial interpolation grid. 

Now, an upper speed bound can be derived that is spatially varying as a function of curvature along the smoothed path trajectory from the LP-result. Using \eqref{eq_dubins}, the radius of curvature can be related to speed as $v=\rho_s \dot{\psi}=\rho_s \frac{g \tan(\phi)}{v}$. The radius of curvature along the smoothed path and maximum roll angle $\phi_\text{max}$ is available. Thus, the upper speed bound can be computed as
\begin{equation}
v_\text{max}(s)= \sqrt{ \rho_s(s) g \tan(\phi_\text{max}) },
\end{equation}
where $s$ represents the spatial coordinate along the smoothed path of length $L$, i.e., $s\in[0,L]$. This permits to formulate our basic control law for speed as,
\begin{equation}
v(s)=\text{min}(v^\text{ref}(s),v_\text{max}(s)),~\forall s\in[0,L].\label{eq_vctrl}
\end{equation}
It implies following a desired reference velocity, and reducing speed as a function of curvature if necessary. Based on \eqref{eq_vctrl}, two refinement steps analogous to \eqref{eq_lgamma}-\eqref{eq_bargamma} are applied to account for rate and limit bounds.

To summarize, the simplicity and compactness of the proposed decoupled approach is highlighted. Both the control of $\gamma$ according to Sect. \ref{subsec_gamma} and the control of $v$ follow algebraic equations. Only the control of roll angel $\phi$ involves an optimization step. However, aided by the spatial modeling approach this is only a small-scale LP and with a hyperparameter-free objective function.

\subsection{Extension for aerobatic flight\label{subsec_aerobatic}}

Above method fails for excessive $z$-coordinate variation when the flight-path angle $\gamma$ approaches 90$^\circ$ and a singularity in \eqref{eq_dubins} occurs.

Without changing the fundamental structure, above method can be adapted to aerobatic flight maneuvers. This can be achieved by 2 steps: a coordinate system transformation and an adaptation of rate and limit bounds.

It is explicitly stressed that this adaptation is \emph{approximative} as will be detailed below. However, it permits to extend planning capabilities to loopings and similar aerobatic maneuvers while maintaining the simple methods from Sect. \ref{subsec_gamma}-\ref{subsec_speedctrl}. 

First, a coordinate system transformation is applied,
\begin{align}
\begin{bmatrix} x_r \\ y_r \\ z_r \end{bmatrix} &= \begin{bmatrix} x \\ z \\ -y \end{bmatrix},\label{eq_xyz_rot}
\end{align}
which implies a 90$^\circ$-rotation around the original $x$-axis, and where the $x_r$, $y_r$  and $z_r$ axes define the new 3D Cartesian coordinate system.

Second, rate and limit bounds are adapted to mimic \emph{in the new coordinate system the dynamics in the old coordinate system}. Therefore, we assign:
\begin{subequations}\label{eq_psi_gamma_r}
\begin{align}
\psi_r &= \gamma,\label{eq_psi_r}\\
\gamma_r &= -\psi,\label{eq_gamma_r}
\end{align}
\end{subequations}
and $\phi_r=\phi-\pi/2$ (since above rotation was carried out around the $x$-axis), and $v_r=v$. The derivative of \eqref{eq_psi_r} yields $g\tan(\phi_r)/v_r=\dot{\gamma}$, from which limit and rate bounds can be derived as
\begin{align}
 \text{arctan}(\frac{-|\dot{\gamma}_\text{min}|v_{\text{min}}}{g}) \leq & ~\phi_r \leq \text{arctan}(\frac{\dot{\gamma}_\text{max}v_{\text{max}}}{g}),\\
-\dot{\phi}_\text{max} \leq & ~\dot{\phi}_r \leq \dot{\phi}_\text{max},
\end{align}
whereby the rate bounds hold since $\phi_r=\phi-\pi/2$. Similarly, 
the derivative of \eqref{eq_gamma_r} yields $\dot{\gamma}_r = -g\tan(\phi)/v$, from which rate bounds can be derived as
\begin{align}
\frac{g \tan(\phi_\text{min})}{v_\text{max}} \leq \dot{\gamma}_r \leq \frac{g \tan(\phi_\text{max})}{v_\text{min}}.
\end{align}
Note that $\gamma_r$ remains unconstrained to permit aerobatic flight such as looping maneuvers.

Lastly, above mimicking step implies a model approximation. Dynamics along the $x$-axis are exact according to \eqref{eq_xyz_rot} and assignment \eqref{eq_psi_gamma_r}, i.e., $v_r\cos(\gamma_r)\cos(\psi_r)=v\cos(\gamma)\cos(\psi)$. In contrast, the mimicking approximation becomes apparent through the dynamics along the $y$-axis (and similarly along the $z$-axis). It holds $v_r\cos(\gamma_r)\sin(\psi_r)\approx v \sin(\gamma)$ only as long as $\cos(\gamma_r)\approx 1$, i.e., for small $\gamma_r$-angles. 

This implys the model approximation is valid for small $\gamma_r$-angles. For planning it encourages (i) to select a dominating plane in the original coordinate system, i.e., either the $(x,y)$-plane or the $(x,z)$-plane, (ii) to plan in that plane, before (iii) re-rotating the computed trajectory back to the original coordinate system if needed. To give an example, for a looping the $(x,z)$-plane would be dominating. In contrast, for a terrain following application the $(x,y)$-plane is dominating. Determining the dominating plane is a heuristic choice, typically relating variation of the reference path along the critical $z$-axis to variation within the $(x,y)$-plane.

\subsection{Extension for tractor path planning along 3D terrain}

In Sect. \ref{subsec_phi} a comment was made how \eqref{eq_3statedynamics} closely resembles the classic kinematic bicycle model for 2D path planning. In fact, the Dubins airplane model \eqref{eq_dubins} can be modified into 
\begin{align}
\begin{bmatrix} \dot{x}\\ \dot{y} \\ \dot{z} \\ \dot{\psi} \end{bmatrix} &= \begin{bmatrix} v \cos(\gamma) \cos(\psi) \\ v \cos(\gamma) \sin(\psi) \\ v \sin(\gamma) \\  \frac{v}{l}\tan(\delta) \end{bmatrix},\label{eq_tractor}
\end{align}
to approximate ground-based path planning accounting for topographically varying terrain, where $\delta$ indicates the steering angle and $l$ the vehicle's wheelbase.

Crucially and in contrast to UAV path planning, $\gamma$ is now not a control variable anymore but instead directly determined by the topography of the terrain where the vehicle is driving.

Furthermore, speed $v$ only occurs in the nominators of \eqref{eq_tractor}. This is in contrast to UAV-modeling. As a result, the analogous derivation of \eqref{eq_spatial} yields a natural elimination of velocity-dependence. Thus, after  spatial transformation, steering control and velocity control are naturally decoupled \cite{plessen2025smoothing}. 

Thus, while the method of Sect. \ref{subsec_gamma}-\ref{subsec_speedctrl} was developed for UAVs, it can be directly adopted for terrain-following 3D path planning applications such as in \cite{plessen20262d}.

\section{Numerical Experiments\label{sec_expts}}

Results of numerical experiments are visualized in Fig. \ref{fig_LP1LP2}, \ref{fig_10ex}, \ref{fig_gamma} and \ref{fig_timespatial}. Additional results are shown in the Appendix.

\begin{figure}[htbp]
\centering
%
\begin{subfigure}[b]{0.24\textwidth}
\centering
\includegraphics[width=1.0\textwidth]{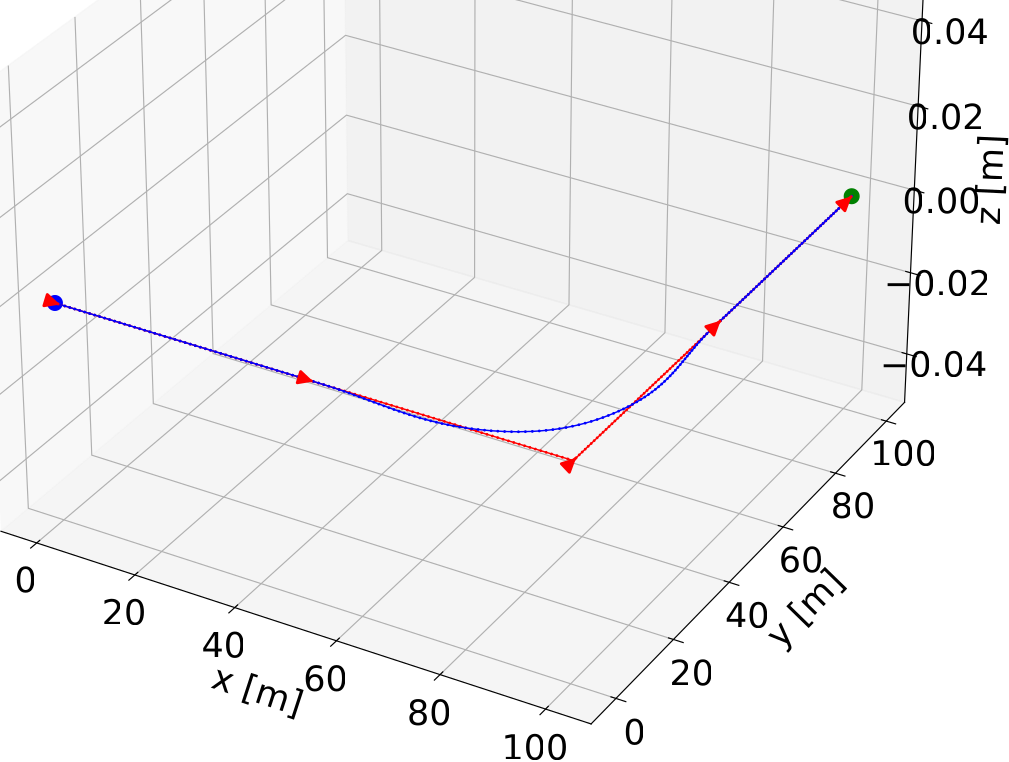}
\caption{\textsf{LP1}. \vspace*{0.1cm}}
\label{fig_LP1LP2_1}
\end{subfigure}
\begin{subfigure}[b]{0.24\textwidth}
\centering
\includegraphics[width=1.0\textwidth]{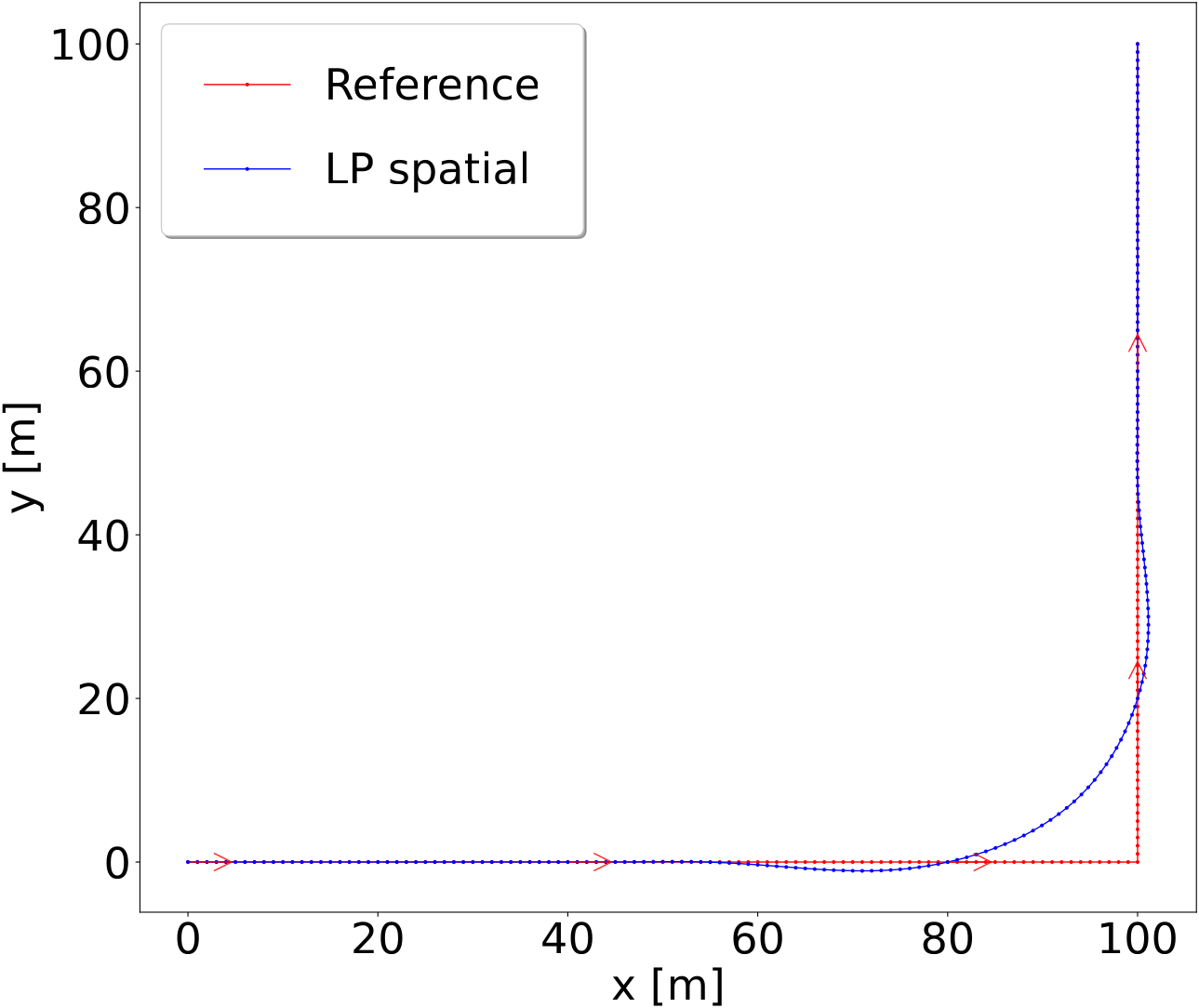}
\caption{$(x,y)$-plane, \textsf{LP1}.\vspace*{0.1cm}}
\label{fig_LP1LP2_2}
\end{subfigure}
%
\begin{subfigure}[b]{0.24\textwidth}
\centering
\includegraphics[width=1.0\textwidth]{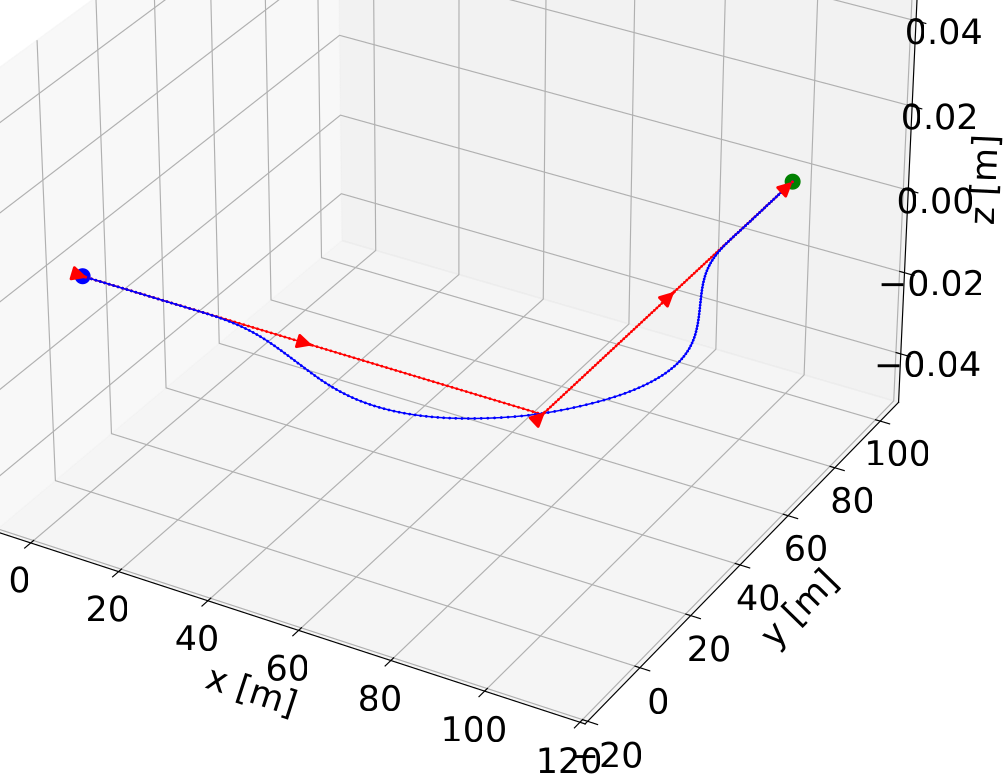}
\caption{\textsf{LP2}. \vspace*{0.1cm}}
\label{fig_LP1LP2_3}
\end{subfigure}
\begin{subfigure}[b]{0.24\textwidth}
\centering
\includegraphics[width=1.0\textwidth]{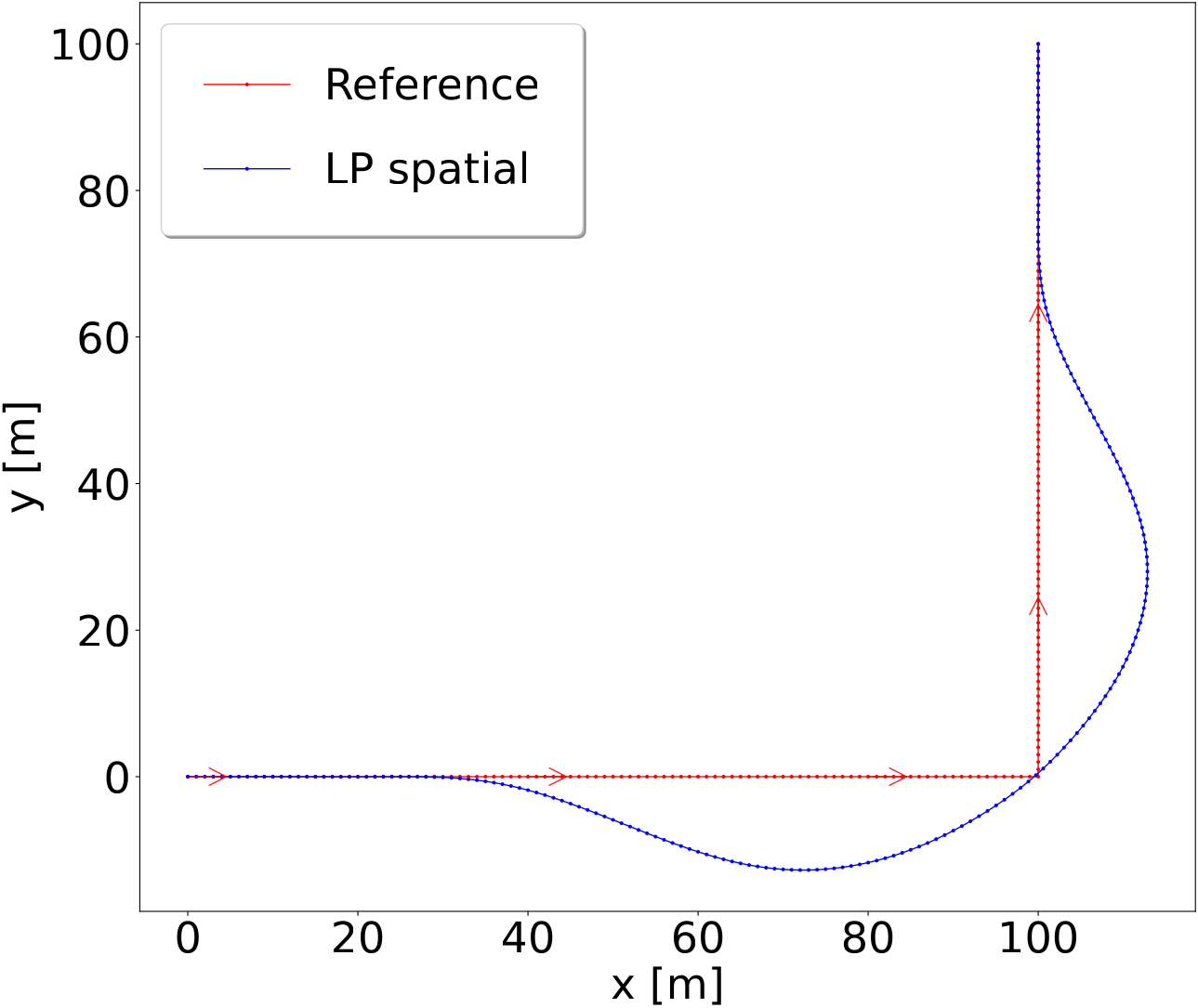}
\caption{$(x,y)$-plane, \textsf{LP2}. \vspace*{0.1cm}}
\label{fig_LP1LP2_4}
\end{subfigure}
%
\caption{Illustration of the effects of \textsf{LP1} and \textsf{LP2} for an edgy reference smoothing example.}
\label{fig_LP1LP2}
\end{figure}

\begin{figure*}[htbp]
\centering
%
\begin{subfigure}[b]{0.19\textwidth}
\centering
\includegraphics[width=1.0\textwidth]{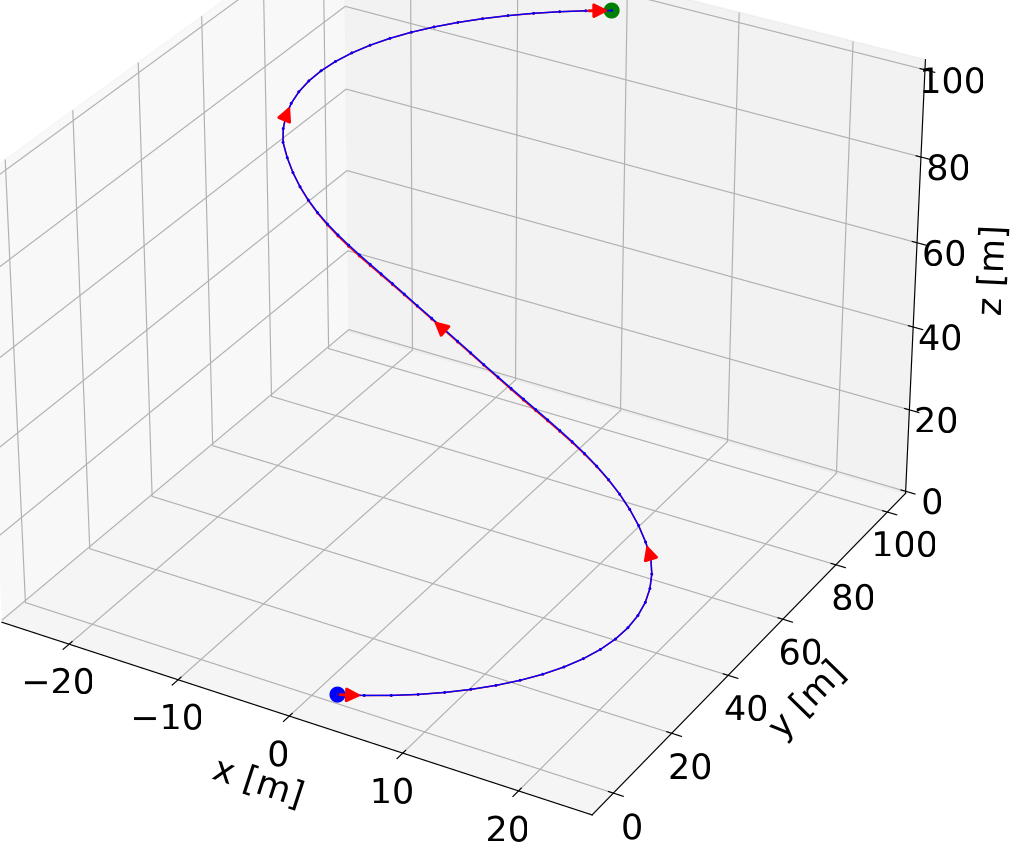}
\caption{Ex. 1, \textsf{LP1}. \vspace*{0.cm}}
\label{fig_cylindr_subf1}
\end{subfigure}
\begin{subfigure}[b]{0.19\textwidth}
\centering
\includegraphics[width=1.0\textwidth]{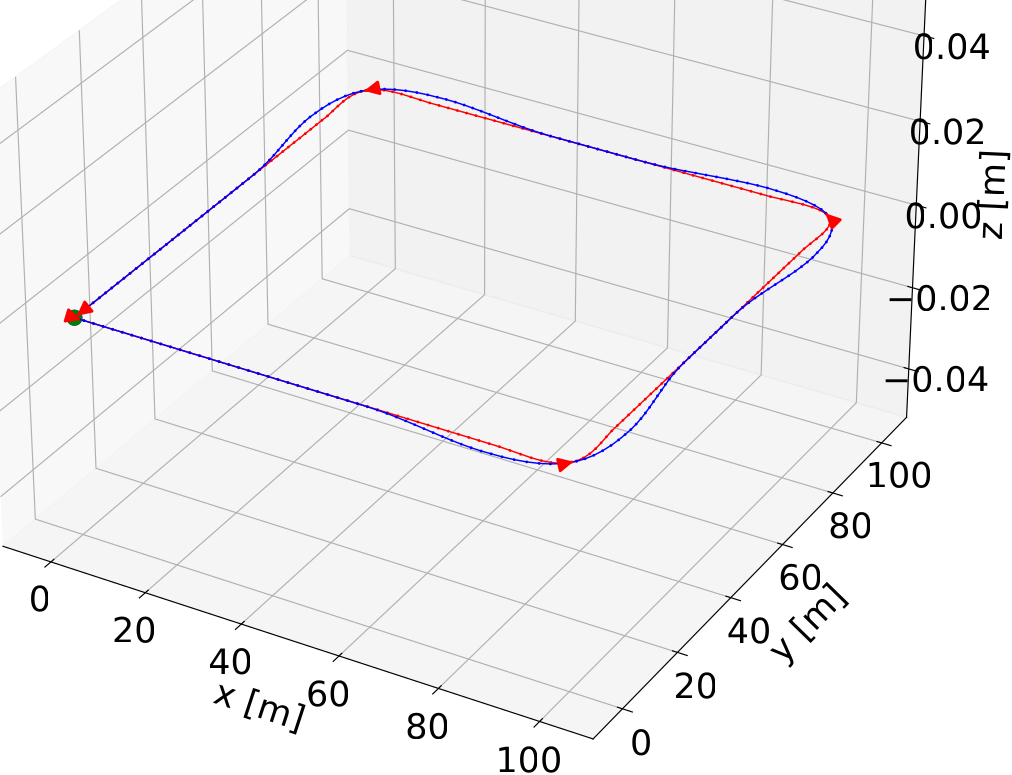}
\caption{Ex. 2, \textsf{LP2}. \vspace*{0.cm}}
\label{fig_cylindr_subf2}
\end{subfigure}
\begin{subfigure}[b]{0.19\textwidth}
\centering
\includegraphics[width=1.0\textwidth]{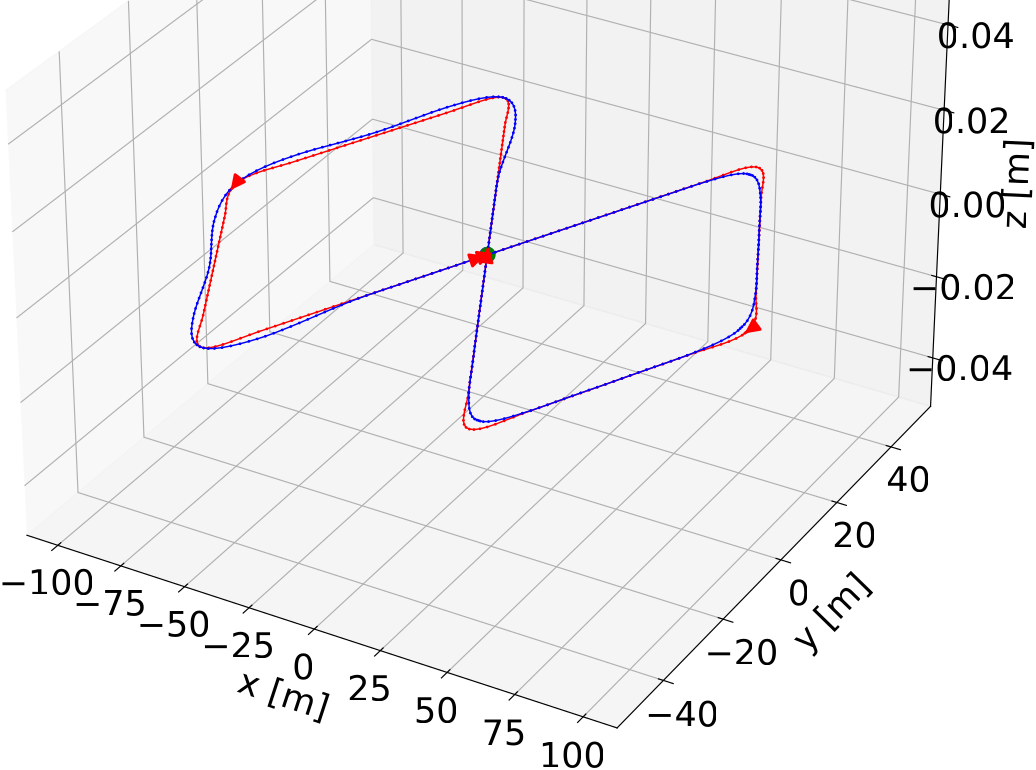}
\caption{Ex. 3, \textsf{LP2}. \vspace*{0.cm}}
\label{fig_cylindr_subf2}
\end{subfigure}
\begin{subfigure}[b]{0.19\textwidth}
\centering
\includegraphics[width=1.0\textwidth]{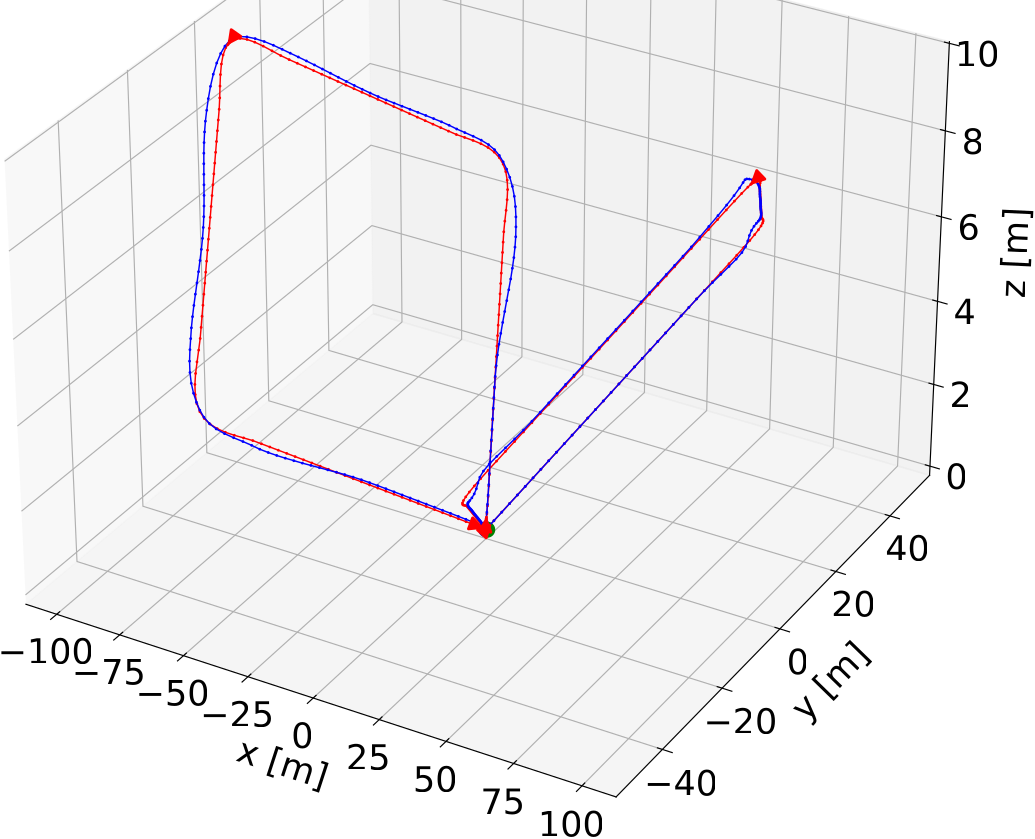}
\caption{Ex. 4, \textsf{LP2}. \vspace*{0.cm}}
\label{fig_cylindr_subf2}
\end{subfigure}
\begin{subfigure}[b]{0.19\textwidth}
\centering
\includegraphics[width=1.0\textwidth]{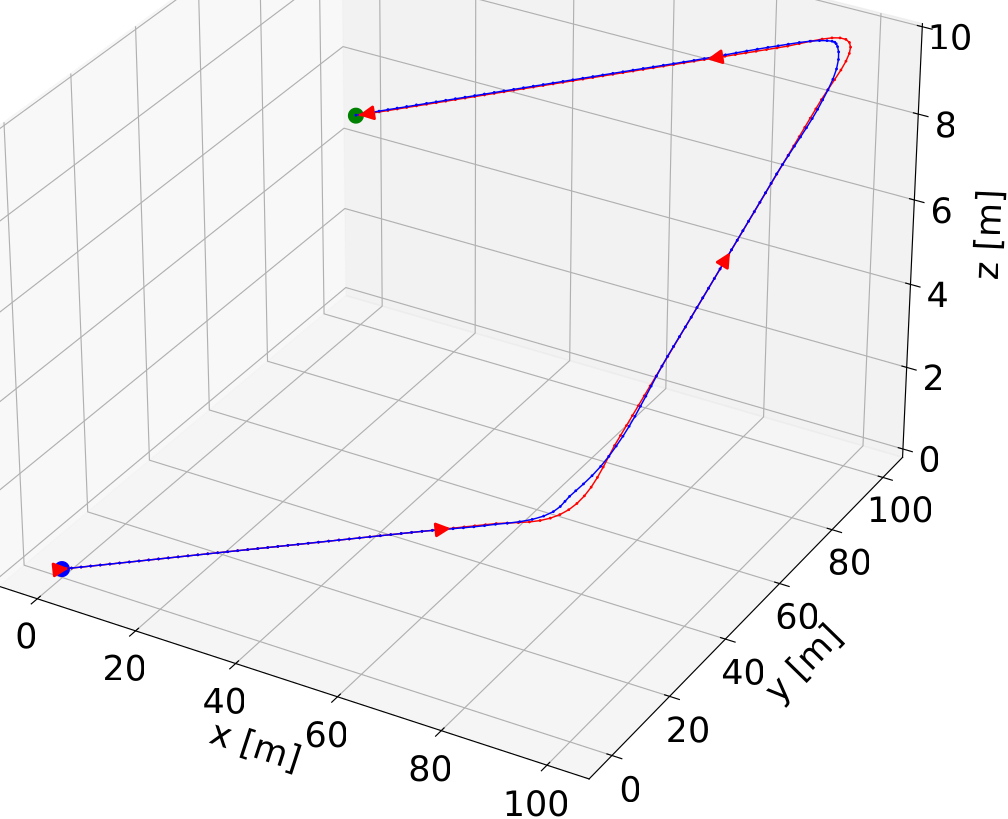}
\caption{Ex. 5, \textsf{LP1}. \vspace*{0.cm}}
\label{fig_cylindr_subf2}
\end{subfigure}
%
\begin{subfigure}[b]{0.19\textwidth}
\centering
\includegraphics[width=1.0\textwidth]{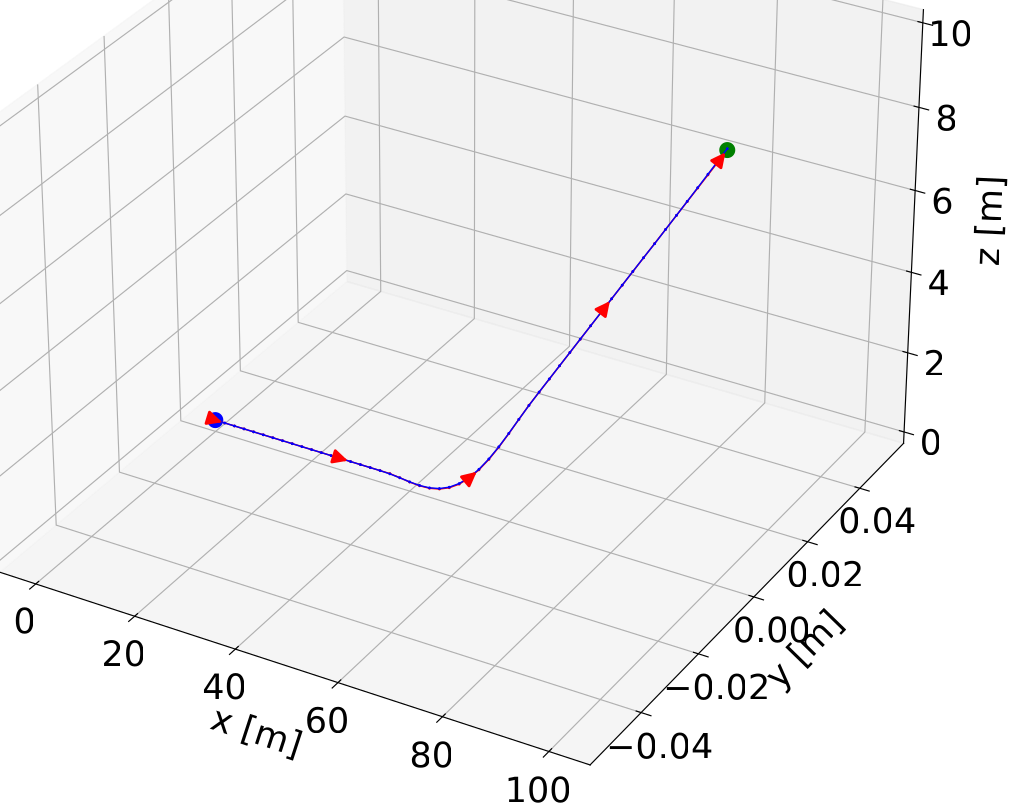}
\caption{Ex. 6, \textsf{LP1}. \vspace*{0.cm}}
\label{fig_cylindr_subf1}
\end{subfigure}
\begin{subfigure}[b]{0.19\textwidth}
\centering
\includegraphics[width=1.0\textwidth]{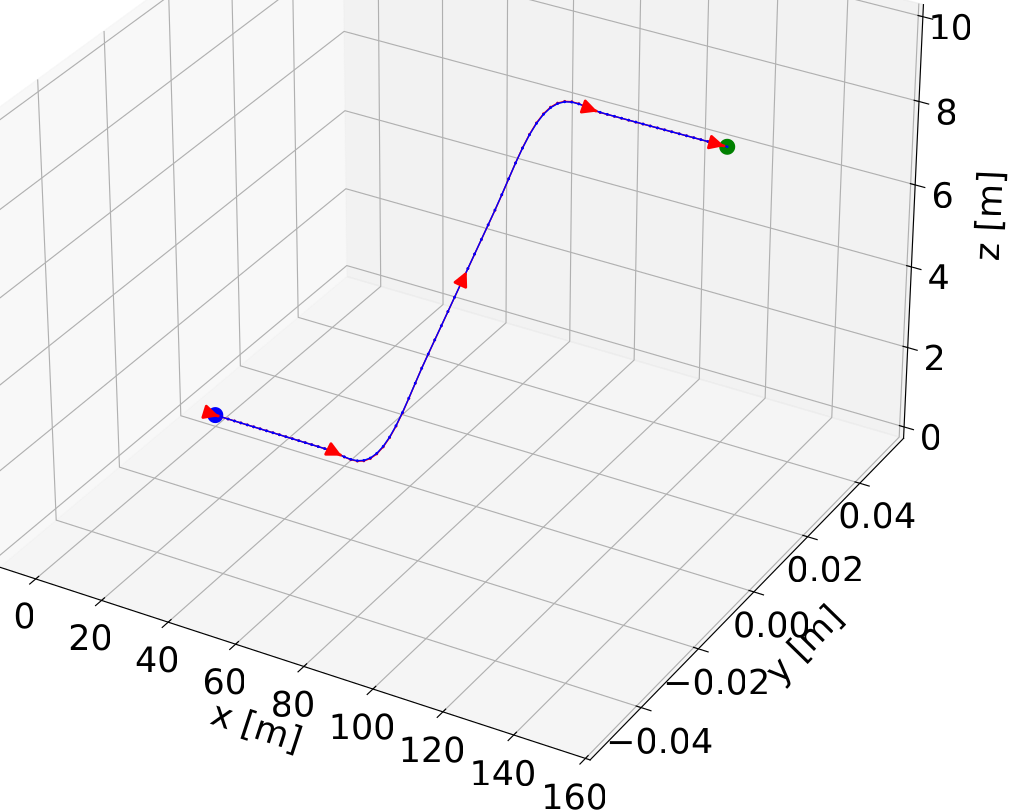}
\caption{Ex. 7, \textsf{LP1}. \vspace*{0.cm}}
\label{fig_cylindr_subf2}
\end{subfigure}
\begin{subfigure}[b]{0.19\textwidth}
\centering
\includegraphics[width=1.0\textwidth]{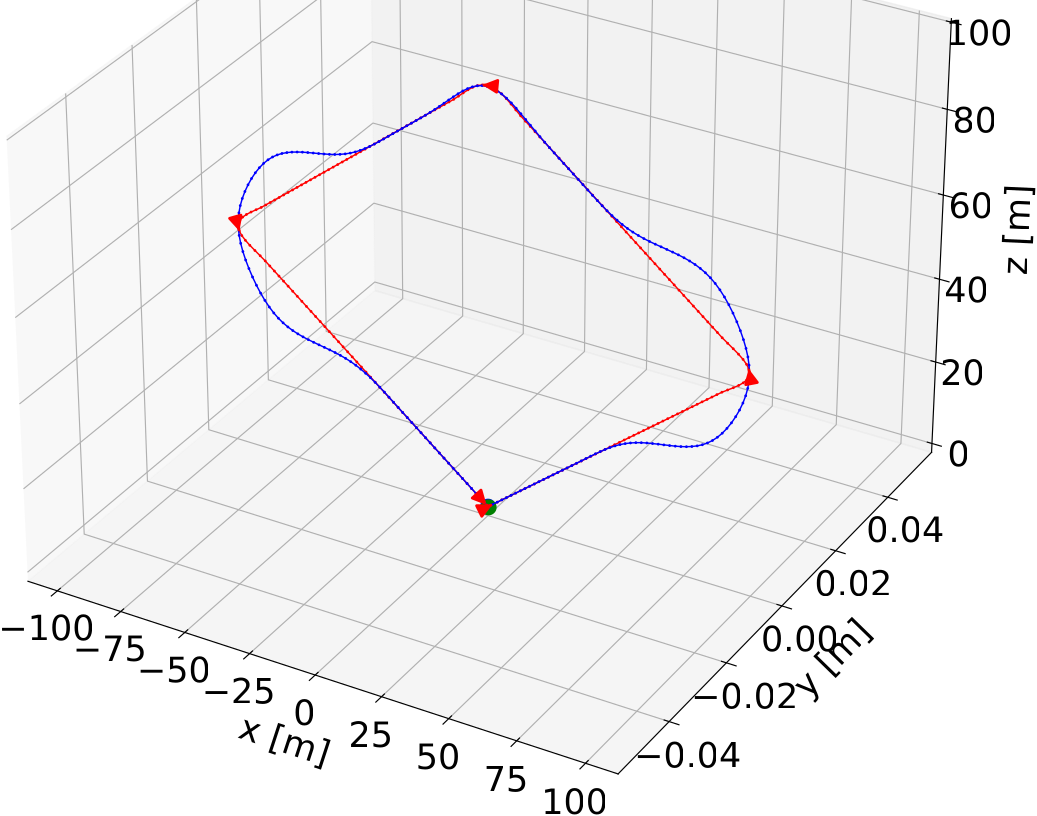}
\caption{Ex. 8, \textsf{LP2}. \vspace*{0.cm}}
\label{fig_cylindr_subf2}
\end{subfigure}
\begin{subfigure}[b]{0.19\textwidth}
\centering
\includegraphics[width=1.0\textwidth]{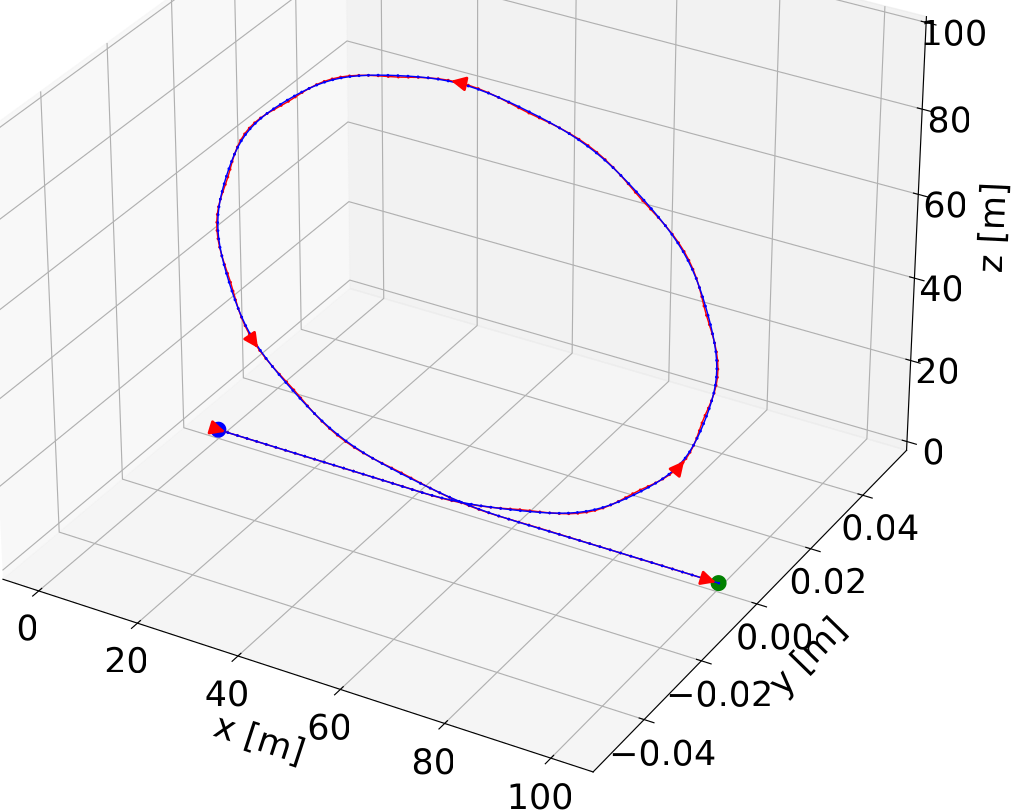}
\caption{Ex. 9, \textsf{LP1}. \vspace*{0.cm}}
\label{fig_cylindr_subf2}
\end{subfigure}
\begin{subfigure}[b]{0.19\textwidth}
\centering
\includegraphics[width=1.0\textwidth]{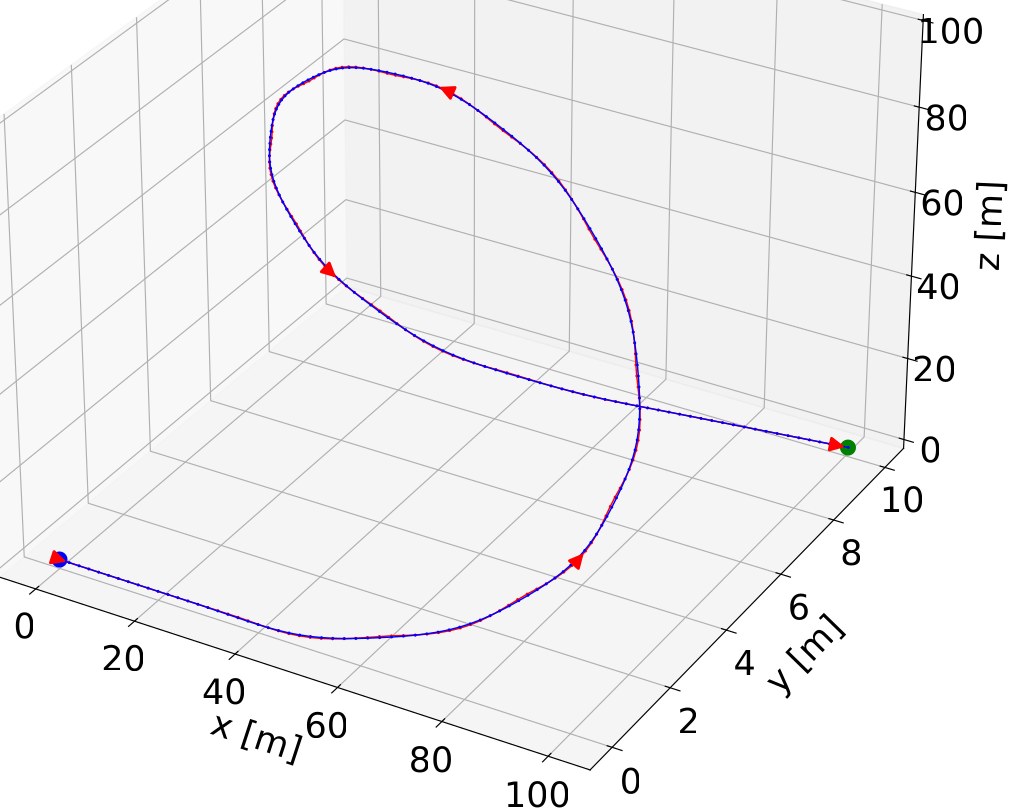}
\caption{Ex. 10, \textsf{LP2}. \vspace*{0.cm}}
\label{fig_cylindr_subf2}
\end{subfigure}
%
\caption{Illustration of 10 numerical experiments. For Ex. 1 a path generated by the method from \cite{mclain2014implementing} served as reference. For the other examples waypoints were defined, then connected by uniform interpolation, before the method from Sect. \ref{subsec_gamma}-\ref{subsec_aerobatic} was applied. For example, for Ex. 3 there were 9 waypoints. }
\label{fig_10ex}
\end{figure*}


\begin{figure}[htbp]
\centering
%
\begin{subfigure}[b]{0.24\textwidth}
\centering
\includegraphics[width=1.0\textwidth]{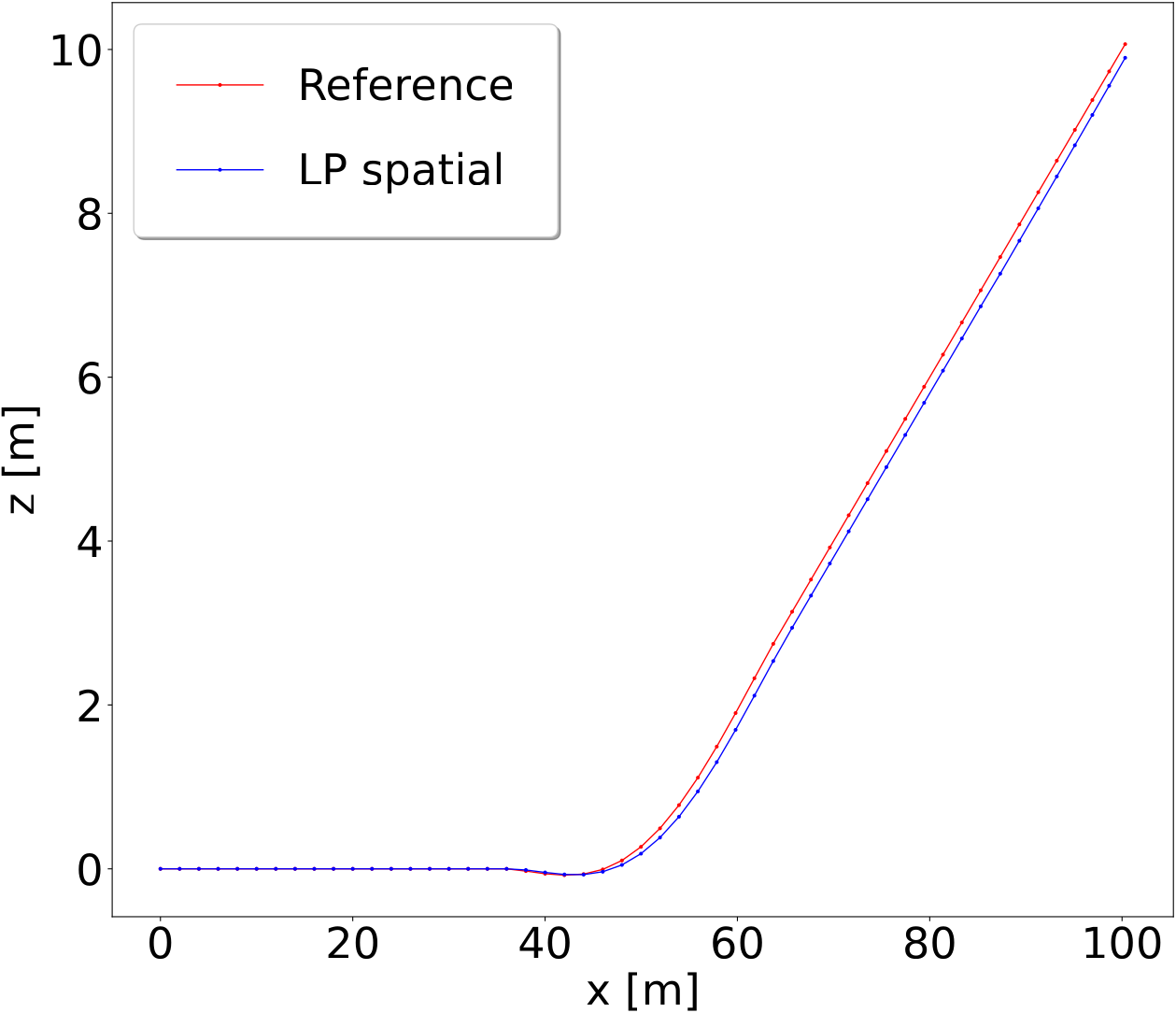}
\caption{Ex. 6, tracking if $\bar{\gamma}_k=\tilde{\gamma}_k$ in \eqref{eq_bargamma}. \vspace*{0.0cm}}
\label{fig_cylindr_subf1}
\end{subfigure}
\begin{subfigure}[b]{0.24\textwidth}
\centering
\includegraphics[width=1.0\textwidth]{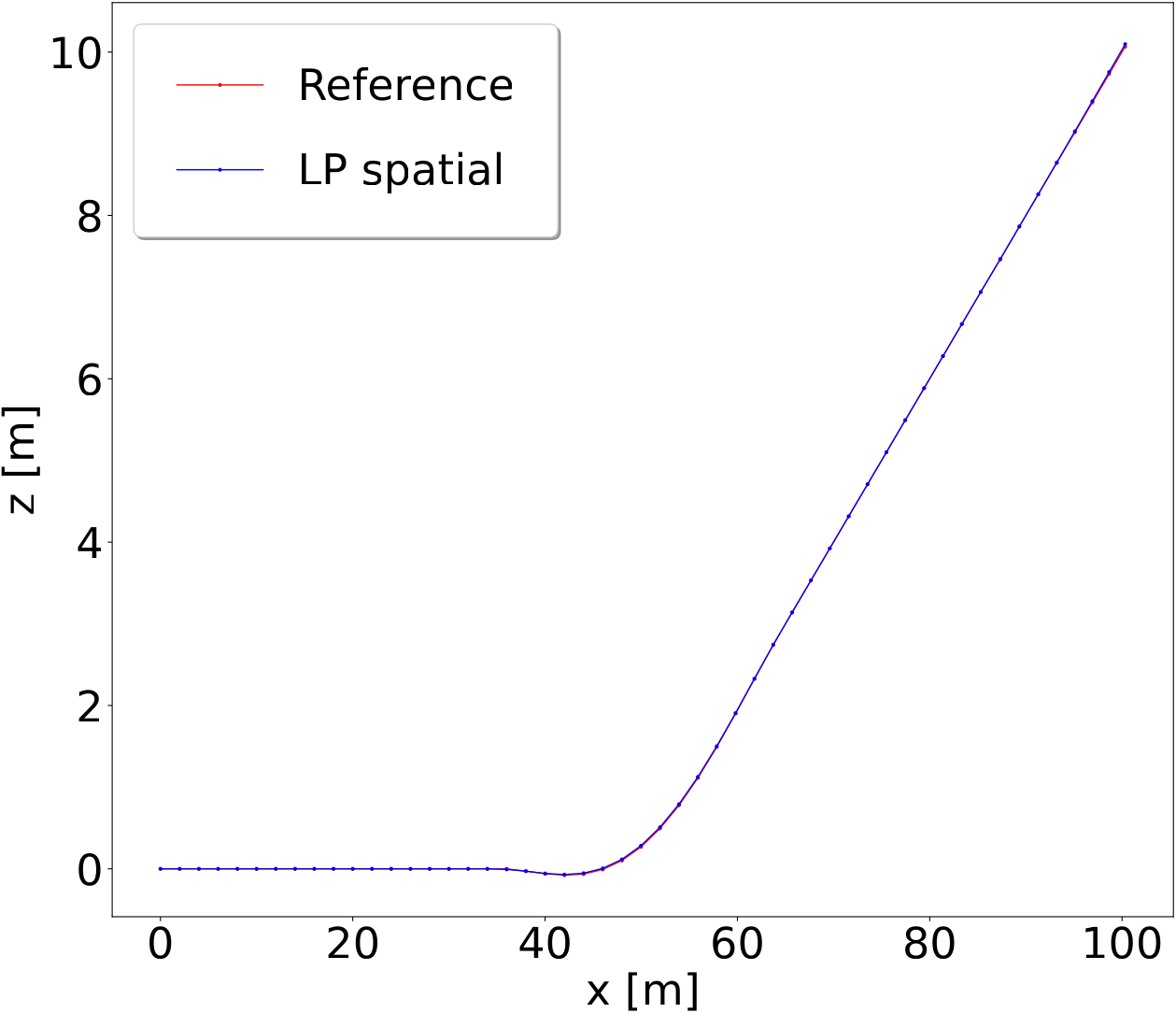}
\caption{Ex. 6, no tracking error for \eqref{eq_bargamma}.\vspace*{0.0cm}}
\label{fig_cylindr_subf2}
\end{subfigure}
%
\caption{Illustration of a detail discussed in Sect. \ref{subsec_gamma}.}
\label{fig_gamma}
\end{figure}

\begin{figure}
\centering
\includegraphics[width=0.8\linewidth]{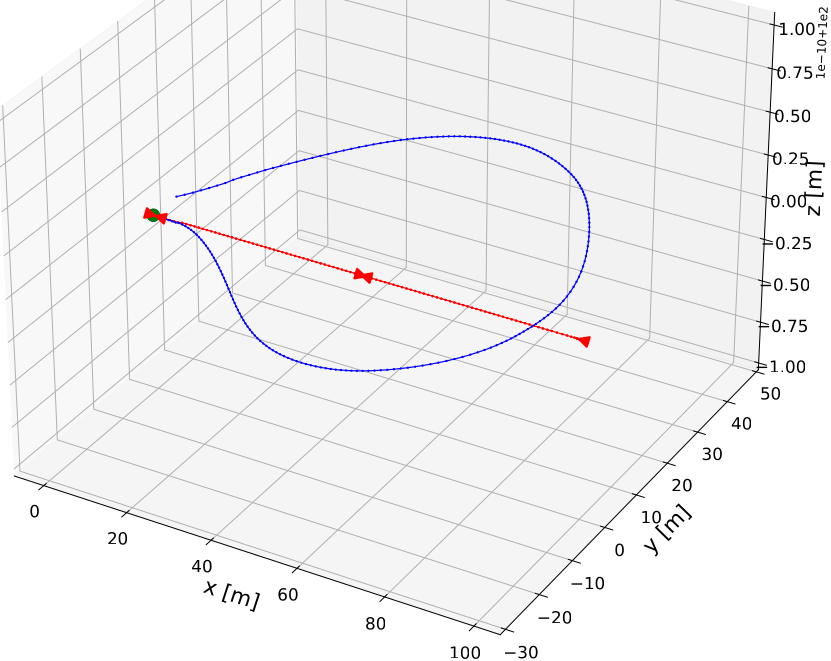}
\caption{A very unfavorable reference can still be tracked. See Sect. \ref{sec_discussion}.}
\label{fig_timespatial}
\end{figure}

\section{Discussion, Limitations and Outlook\label{sec_discussion}}

The \emph{Dubins airplane model} is a simplifying model \cite{mclain2014implementing}. Limitations include (i) not accounting for wind influences, (ii) time-delays in inner low-level control loops that regulate flight-path angle, roll angle and speed, and (iii) aerodynamic effects such as, for example, a positive flight-path angle reducing airspeed.

While objective functions of the LPs in Sect. \ref{subsec_phi} are hyperparameter free, there are still heuristic choices needed. These are (i) path tangential interpolation spacing, and (ii) in case of aerobatic maneuvers a heuristic to determine the dominating plane and a splitting method to determine when to start planning in the respective coordinate systems.

The title of the paper deliberately includes the phrase trajectory \emph{smoothing}. This is to underline that a reasonable reference is required as input. Not any arbitrary references can be smoothed. For example, a 3-waypoint reference with $\{(x_i^\text{ref},y_i^\text{ref},z_i^\text{ref})\}=\{(0,0,100),(100,0,100),(0,0,100)\}$, that implies an instantaneous 180$^\circ$-turn, which is infeasible for the nonholonomic dynamics of the Dubins airplane model, cannot directly be smoothed with the presented method. This is because spatial modeling implies a parametrization with respect to a reference trajectory, which makes it vulnerable to singularities \cite{arrizabalaga2024universal,fork2023euclidean}. One practical simple solution is to (i) linearize and discretize \eqref{eq_dubins} in the time domain and solve an analogous tracking problem as \eqref{eq_LP1} for above 3-waypoint reference, before (ii) solving e.g. \eqref{eq_LP2} on top to add spatial shaping constraints. This approach is efficient since optimization problems based on discretized time domain dynamics of \eqref{eq_dubins} can be solved robustly (e.g., nilpotent state transition matrices). In contrast, formulating spatial constraints such as in \eqref{eq_LP2} is not obvious for the time domain.

The speed control law \eqref{eq_vctrl} encourages following of a reference velocity as long as path curvature permits this. Alternatively, the method from \cite{plessen2017trajectory} may be adopted for \emph{time-scheduling of the reaching of waypoints} along the path. In that case, in contrast to the method of Sect. \ref{subsec_phi}, two alternative LPs would be formulated with now 2 controls, $u=[\frac{1}{v},\phi]$, thereby replacing the speed control method from Sect. \ref{subsec_speedctrl}. 

A perceived benefit of Dubins airplane model-based planning  is that 6 variables are involved that are also typical for quadrotor path planning. These are the 3 position coordinates $(x,y,z)$ and 3 angles $(\psi,\gamma,\phi)$. Dynamical quadrotor models typically have a much larger state space. Future work may analyze to what extent the fast (to a large part algebraic) planning methods presented may be used for reference trajectory generations for quadrotors \cite{lin2014path}. It is interesting to note that the 4 flat outputs characteristic for \emph{differential flatness}-based algorithms \cite{mellinger2011minimum} are the exact same as the 4 states of the Dubins airplane model in \eqref{eq_dubins}.

Finally, spatial modeling seems well suited for drone formation flights \cite{wang2019efficient, chao2012uav}. Setups can be imagined where a leading drone is followed by a cascade of follower drones. For such setups the spatial modeling of Sect. \ref{subsec_phi} could be extended with longitudinal (adaptive cruise control-like) dynamics \cite{plessen2016multi}.

\section{Conclusion\label{sec_conclusion}}

A method for trajectory smoothing for UAV reference path planning was presented. It was derived based on the dynamics of a Dubins airplane model, and involved a decoupling step, spatial modeling and linear programming. 3 control variables were considered. Proposed decoupling method resulted in algebraic control laws for flight-path angle and speed control. Only for roll angle control an optimization step was required, involving the solution of a small linear program. Two variations were discussed, differing by reference centerline tracking and the introduction of a path shaping constraint. An extension to aerobatic flight was discussed, which came at the cost of a model approximation, however at the gain of maintaining the general model structure. An extension to tractor path planning along 3D terrain was discussed. 

It was found that despite its simplicity proposed method could be applied to a variety of 3D path planning setups.

%
%
%
%
%

\nocite{*}

\bibliographystyle{IEEEtran}

\bibliography{myref}



\appendix


\begin{figure*}[htbp]
\centering
%
\begin{subfigure}[b]{0.49\textwidth}
\centering
\includegraphics[width=1.0\textwidth]{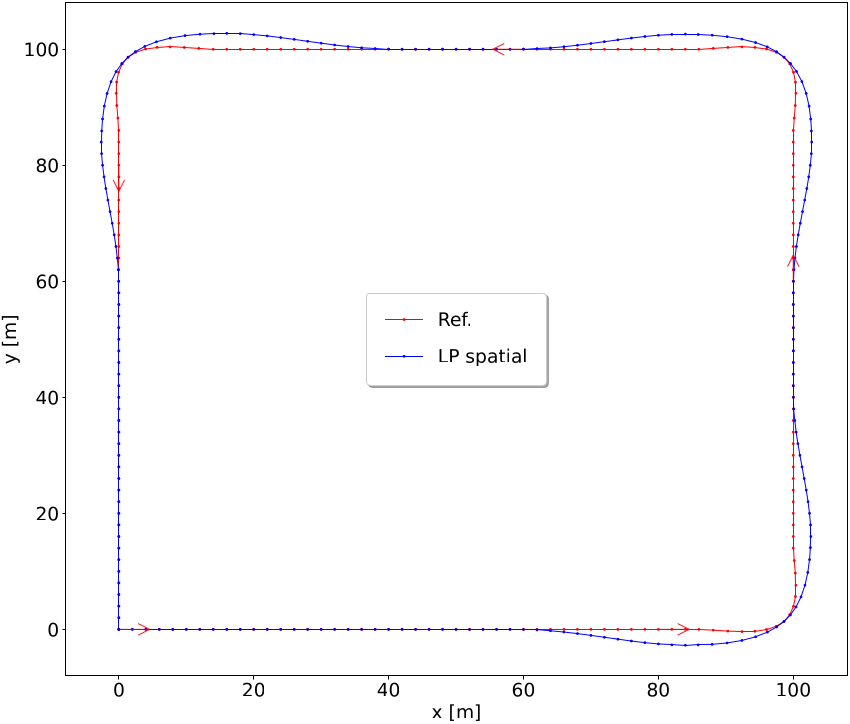}
\caption{Ex. 2, \textsf{LP2}. \vspace*{0.3cm}}
\label{fig_planeview_1}
\end{subfigure}
\begin{subfigure}[b]{0.49\textwidth}
\centering
\includegraphics[width=1.0\textwidth]{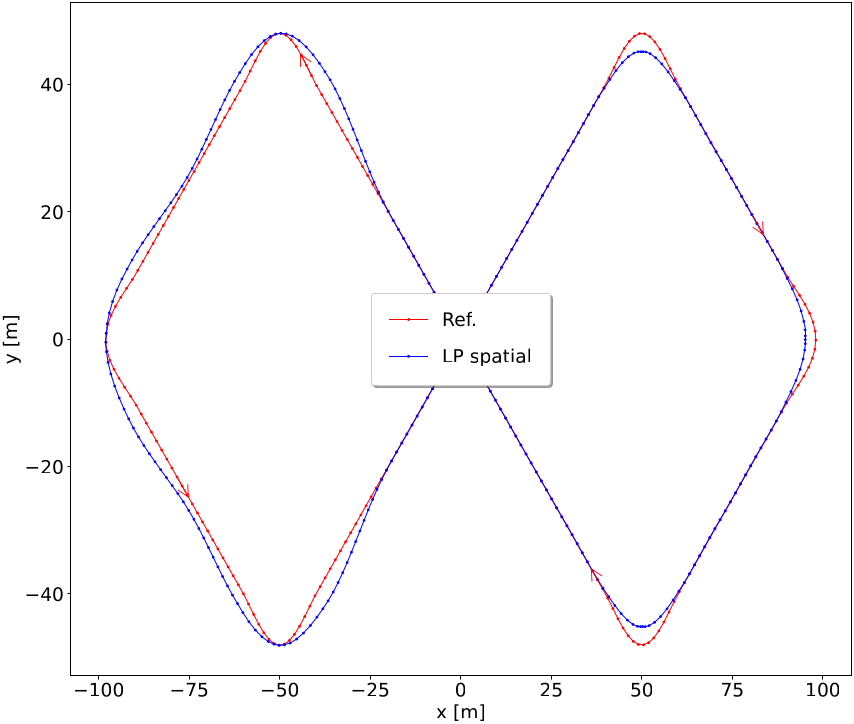}
\caption{Ex. 4, \textsf{LP2}. Note how the resulting path stays to the right-hand side of the reference. This is in line with the formulation of \textsf{LP2} in \eqref{eq_LP2}. \vspace*{0.3cm}}
\label{fig_planeview_2}
\end{subfigure}
\begin{subfigure}[b]{0.49\textwidth}
\centering
\includegraphics[width=1.0\textwidth]{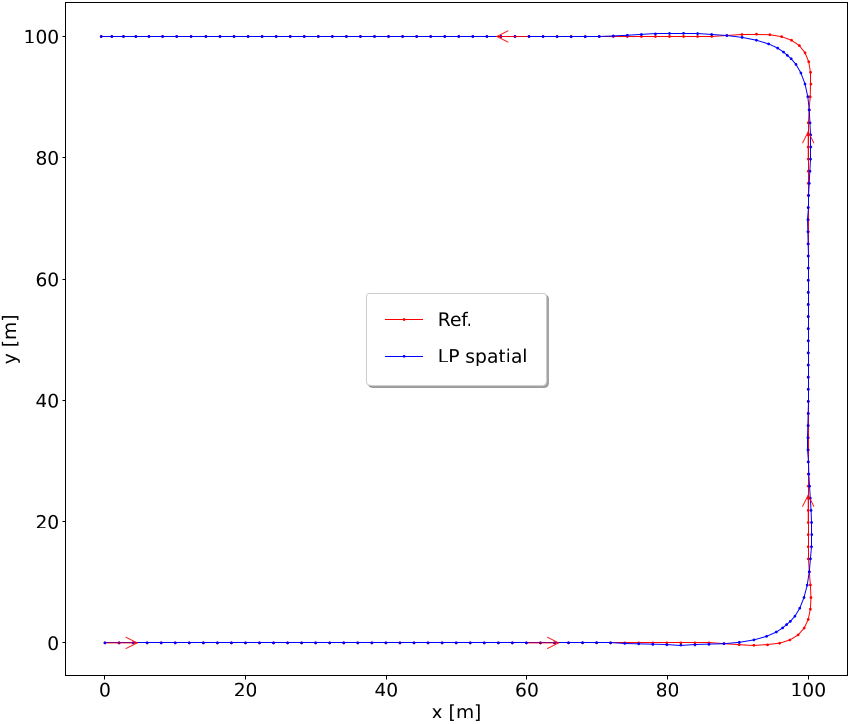}
\caption{Ex. 5, \textsf{LP1}. \vspace*{0.3cm}}
\label{fig_planeview_3}
\end{subfigure}
\begin{subfigure}[b]{0.49\textwidth}
\centering
\includegraphics[width=1.0\textwidth]{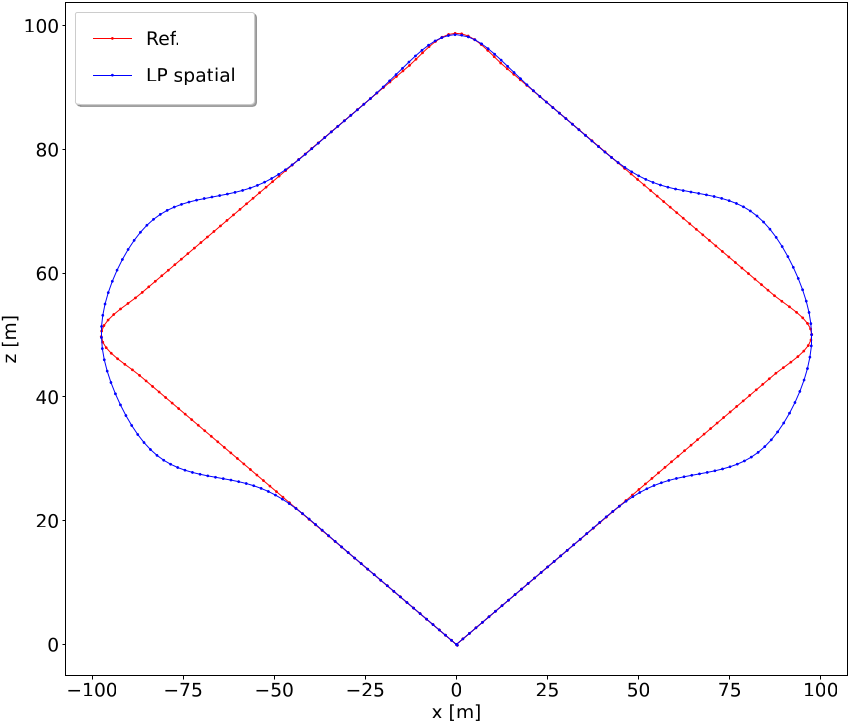}
\caption{Ex. 8, \textsf{LP2}. \vspace*{0.3cm}}
\label{fig_planeview_4}
\end{subfigure}
\begin{subfigure}[b]{0.49\textwidth}
\centering
\includegraphics[width=1.0\textwidth]{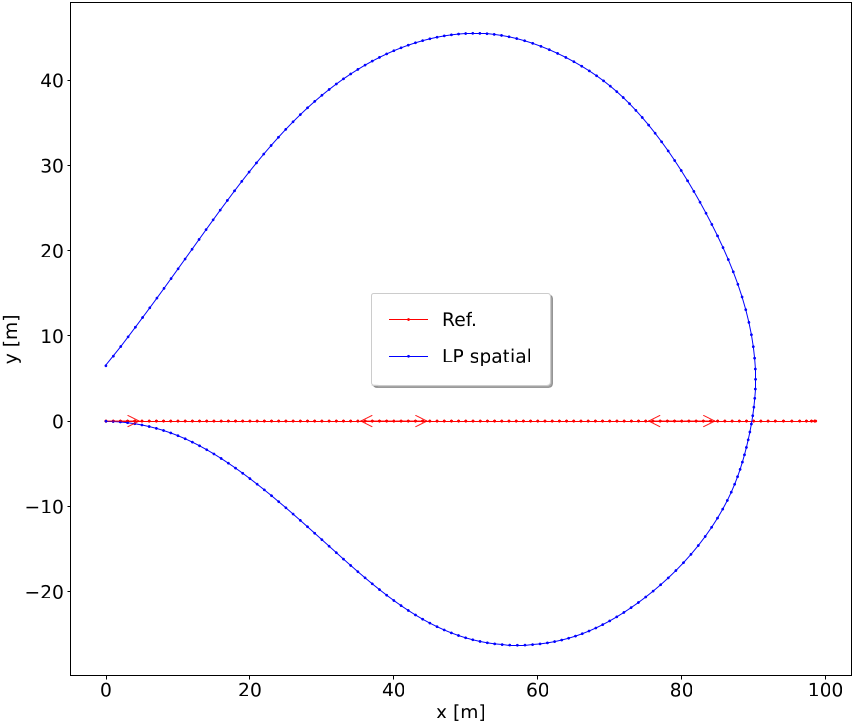}
\caption{Example of Fig. \ref{fig_timespatial}, \textsf{LP1}. \vspace*{0.3cm}}
\label{fig_planeview_5}
\end{subfigure}
%
\caption{Detailed view of examples in Fig. \ref{fig_10ex} and Fig. \ref{fig_timespatial} in the $(x,y)$- or $(x,z)$-plane.}
\label{fig_planeview}
\end{figure*}

\end{document}